\documentclass{amsart}

\theoremstyle{definition}

\theoremstyle{remark}

\numberwithin{equation}{section}

\begin{document}

\title{Polytope sums and Lie characters}

\author{Mark A. Walton}
\address{Department of Physics, University of Lethbridge, 
Lethbridge, Alberta, Canada T1K 3M4}
\email{walton@uleth.ca}
\thanks{This work was supported in part by NSERC}



\date{November, 2002}

\dedicatory{This paper is dedicated to the late Professor R. T. Sharp.}


\begin{abstract} 
A new application of polytope theory to Lie theory is presented.   
Exponential sums of convex lattice polytopes are applied 
to the characters of irreducible representations of simple Lie algebras. 
The Brion formula is used to write a polytope expansion of 
a Lie character, that 
makes more transparent certain degeneracies of weight-multiplicities beyond 
those explained by Weyl symmetry. 
\end{abstract}

\maketitle



\section{Introduction} 

A polytope is the convex hull of finitely many points in $\mathbb R^d$ 
(see \cite{Zi}, for example). A lattice polytope (also known as an integral 
polytope) is a polytope with all 
its vertices in an integral lattice $\Lambda\in \mathbb R^d$ 
(see \cite{B}, for example). The 
exponential sum of a lattice polytope $Pt$, 
\begin{equation}
\sum_{x\in Pt\cap \Lambda}\, \exp\{\langle c, x\rangle\}\ ,
\end{equation}
is a useful tool. Here $c$ is a vector in $\mathbb R^d$, and 
$\langle\cdot,\cdot\rangle$ is the usual inner product. 
For simplicity of notation, we'll 
consider {\it formal} exponential sums
\begin{equation}
E[Pt; \Lambda]\ :=\ \sum_{x\in Pt\cap\Lambda }\, e^x\ .
\end{equation}
Here the formal exponential $e^x$ satisfies 
\begin{equation}
e^xe^y\ =\ e^{x+y}\ ,
\end{equation}
for all $x,y\in \Lambda $, and simply stands for the function 
$e^x(c):= e^{\langle c,x\rangle}$. These exponential sums are the 
generating functions for the integral points in a lattice polytope. 

\vskip10pt
\begin{quotation}
As an extremely simple example, consider the one-dimensional lattice 
polytope with vertices (2) and (7). Its exponential sum is 
\begin{equation}
e^{(7)}+e^{(6)}+e^{(5)}+e^{(4)}+e^{(3)}+e^{(2)}\ . 
\end{equation}

As a very simple example, consider the lattice polytope in 
$\mathbb Z^2$ with vertices $(0,0),(1,0)$ and $(1,1)$. Its 
exponential sum is just 
\begin{equation}
e^{(1,1)}+e^{(1,0)}+e^{(0,0)}\ ,
\end{equation}
since its vertices are the only lattice points it contains. 
\end{quotation}
\vskip10pt

We will apply knowledge of these polytope sums to the calculation of 
the characters of representations of simple Lie algebras (Lie 
characters, for short). Polytope theory and Lie theory may have much to 
teach each other, and we hope this contribution will prompt others.   

This work is preliminary. Proofs are not given, and we concentrate on 
low-rank examples. A fuller treatment will be given in \cite{GW}. 

We are grateful to Chris Cummins, Terry Gannon and Pierre Mathieu 
for comments on the manuscript. 
 
\vskip20pt
\section{Brion formula}

Brion \cite{Br} has proved a formula for the exponential sum of a convex 
lattice polytope, that expresses it as a sum of simpler terms, 
associated with each of the vertices of the polytope. It reads
\begin{equation}
E[Pt; \Lambda ]\ =\ \sum_{x\in Pt\cap\Lambda}\, e^x\ =\ 
\sum_{v\in {\rm Vert}Pt}\, e^v\, \sigma_v\ .
\end{equation}
Here ${\rm Vert}Pt$ is the set of vertices of $Pt$. 

The vertex term $\sigma_v$ is determined by the 
cone $K_v$, defined by extending the vertex $v\in {\rm Vert}Pt$, 
\begin{equation}
K_v\ =\ \{\,x\,|\, v+\epsilon x\in Pt,\ {\rm for\ all\ sufficiently\ 
small\ }\epsilon >0\,\}\ .
\end{equation}  
In other words, $K_v$ is generated by the 
vectors $u_i = v_i-v$, where $[v_i,v]$ is an edge of $Pt$, that 
indicate the feasible directions at $v$; it is 
the cone of feasible directions at $v$.\footnote{\,
We will assume that 
the vectors $u_i$ are linearly independent for all 
$v\in {\rm Vert}Pt$, i.e., that all the $K_v$ are 
simple cones, and so $Pt$ is a simple polytope.}  

$\sigma_v$ is associated 
with the exponential sum of $K_v$. Precisely, if $K_v$ is generated 
by vectors $u_1,\ldots,u_k\in \Lambda$, then the series 
$\sum_{x\in K_v\cap\Lambda} \exp\{\langle c,x\rangle\}$ converges for 
any $c$ such that $\langle c,u_i\rangle <0$, for all $i=1,\ldots,k$. 
The series defines a meromorphic function of $c$, $\sigma_v(c)$
\cite{B}.

A cone $K_v$ is unimodular if the fundamental parallelopiped bounded by 
its feasible directions $u_i$ contains no lattice points 
in its interior. If $K_v$ is unimodular, its exponential sum takes the 
very simple form of a multiple geometric series
\begin{equation}
\sigma_v\ =\ \prod_{i=1}^d\, \left(\, 1-e^{u_i}\,\right)^{-1}\ .
\end{equation}
$Pt$ is totally unimodular if $K_v$ is unimodular for all 
$v\in {\rm Vert}Pt$. If a cone is not unimodular, its exponential 
sum is not so simple, but is still easy to write; the 
fundamental parallelopiped of a non-unimodular cone is always a finite union 
of unimodular parallelopipeds. 

\vskip10pt
\begin{quotation}
For the extremely simple one-dimensional example considered above, 
the Brion formula gives 
\begin{equation}
\frac{e^{(7)}}{1-e^{(-1)}}\ +\ \frac{e^{(2)}}{1-e^{(1)}}\ .   
\end{equation}
If we use 
\begin{equation}
\frac{1}{1-e^{x}}\ =\ \frac{-e^{-x}}{1-e^{-x}}\ ,
\end{equation}
this becomes 
\begin{equation}
\frac{e^{(7)}-e^{(1)}}{1-e^{-(1)}}\ ,   
\end{equation}
i.e., the simple geometric series, so that (1.4) is recovered. 

For the $d=2$ example above, the Brion formula (2.1),(2.3) gives 
\begin{equation}
\begin{split}
\frac{e^{(1,1)}}{(1-e^{(-1,-1)})(1-e^{(0,-1)})}\ 
\ \  &+\ \ \   
\frac{e^{(1,0)}}{(1-e^{(-1,0)})(1-e^{(0,1)})}\\  +\ \ \   &  
\frac{e^{(0,0)}}{(1-e^{(1,0)})(1-e^{(1,1)})}\ . 
\end{split}
\end{equation}
This simplifies to (1.5). 
\end{quotation}
\vskip10pt

\vskip20pt
\section{Weyl character formula}

Let $P = \mathbb Z\{\Lambda^i\,|\, i=1,\ldots,r \}$ 
denote the weight lattice of a simple Lie algebra $X_r$, of rank 
$r$. Here $\Lambda^i$ stands for the $i$-th fundamental weight. 
$R_>$ ($R_<$) will denote the set of positive (negative) roots 
of $X_r$, and 
$S=\{\alpha_i\,|\, i=1,\ldots,r \}$ its simple roots.  

The highest weights of integrable irreducible representations of 
$X_r$ belong to the set $P_\ge = \{\lambda = \sum_{j=1}^r\lambda_j\Lambda^j 
\,|\, \lambda_j\in \mathbb Z_{\ge 0} \}$. 
The formal character of the irreducible representation $L(\lambda)$ 
of highest weight 
$\lambda$ is 
\begin{equation}
{\rm ch}_\lambda\ =\ \sum_{\mu\in P}\, {\rm mult}_\lambda(\mu)\, e^\mu\ ,
\end{equation} 
where ${\rm mult}_\lambda(\mu)$ denotes the multiplicity of weight $\mu$ 
in $L(\lambda)$. 
The famous Weyl character formula is 
\begin{equation}
{\rm ch}_\lambda\ =\ 
\frac{\sum_{w\in W}\, (\det w)\, e^{w.\lambda}}
{\prod_{\alpha\in R_>}\, (1-e^{-\alpha})}\ .
\end{equation}
Here $W = \langle r_i\,|\, i=1,\ldots,r \rangle$ is the Weyl group of $X_r$, 
and $r_i$ is the $i$-th 
primitive Weyl reflection, with action $r_i\mu = 
\mu - \mu_i\alpha_i$ on $\mu\in P$. $\det w$ is the sign of $w\in W$, and 
$w.\lambda = w(\lambda+\rho)-\rho$ is the shifted action of $w$ on $\lambda$, 
with the Weyl vector $\rho = \sum_{i=1}^r \Lambda^i = 
\sum_{\alpha\in R_>} \alpha/2$. 

After using ${\rm ch}_0 = 1$ to derive the Weyl denominator formula, we can 
rewrite the character formula as 
\begin{equation}
{\rm ch}_\lambda\ =\ 
\frac{\sum_{w\in W}\, (\det w)\, e^{w(\lambda+\rho)}}
{\sum_{w\in W}\, (\det w)\, e^{w\rho}}\ .
\end{equation} 
Comparing to (3.1) reveals the Weyl symmetry 
\begin{equation}
{\rm mult}_\lambda(\mu)\ =\ {\rm mult}_\lambda(w\mu)\ \ \ \ (\forall w\in W) 
\end{equation}
of the weight multiplicities. 

Alternatively, we can rewrite (3.2) in a different form that is  
also manifestly $W$ symmetric:
\begin{equation}
{\rm ch}_\lambda\ =\ 
\sum_{w\in W}\, e^{w\lambda}\, \prod_{\alpha\in R_>}\, 
(1-e^{-w\alpha})^{-1}\ . 
\end{equation} 
The usual formula (3.2) is recovered from this using (2.5). Each 
Weyl element $w\in W$ separates the 
positive roots into two disjoint sets:
\begin{equation}
\begin{split}
R_>^w:= \{\alpha\in R_>\,|\, w\alpha\in R_>\}\ ,&\ \ 
R_<^w:= \{\alpha\in R_>\,|\, w\alpha\in R_<\}\ ,\\ 
R_>^w \cup R_<^w\ =\ R_>\ ,&\ \ R_>^w \cap R_<^w\ =\ \{\}\ ,\\ 
wR_>^w\ =\ R_>^w\ \ \ \ ,&\ \ \ \ wR_<^w\ =\ -R^w_<\ \  .
\end{split}
\end{equation}
It can be shown that $\det w = (-1)^{\Vert R^w_<\Vert}$, and 
\begin{equation}
-w\rho +\rho\ =\ \sum_{\beta\in R^w_<}\, \beta\ \ .
\end{equation}
Using these results, (3.5) becomes 
\begin{equation}
\begin{split}
{\rm ch}_\lambda\ =&\ \sum_{w\in W}\, e^{w\lambda}\, 
\prod_{\beta\in R^w_<}\, (-e^{w\beta})(1-e^{w\beta})^{-1}\, 
\prod_{\alpha\in R^w_>}\, (1-e^{-w\alpha})^{-1}\ \\ 
=&\ \sum_{w\in W}\,(\det w)\, e^{w\lambda-w\sum_{\gamma\in R^w_<}\gamma}\, 
\prod_{\beta\in R^w_<}\, (1-e^{w\beta})^{-1}\, 
\prod_{\alpha\in R^w_>}\, (1-e^{-w\alpha})^{-1}\ ,
\end{split}
\end{equation}
so that (3.2) follows.

\vskip20pt
\section{Character polytope-expansion}

The form (3.5) of the Weyl character formula is 
similar to the Brion formula. To make this more precise, we'll write 
the Brion formula for the exponential sum of 
the lattice polytope $Pt_\lambda$ with vertices in the 
Weyl orbit of a highest weight $\lambda\in P_\ge$, i.e. 
${\rm Vert}Pt_\lambda = 
W\lambda$. The appropriate lattice here is the root lattice $Q$ of $X_r$,  
shifted by $\lambda$: $\lambda + Q\subset P$. If we define
\begin{equation}
P(\lambda)\ :=\ \{\, \mu \,|\, {\rm mult}_\lambda(\mu)\not=\  0\,  \,\}\ 
\subset\  P\, ,
\end{equation}
then
\begin{equation}
Pt_\lambda\cap (\lambda+Q)\ =\ P(\lambda)\ ,
\end{equation} 
and the polytope sum will be 
\begin{equation}
E[Pt_\lambda; \lambda+Q]\ =\ \sum_{\mu\in P(\lambda)}\, e^\mu\ .
\end{equation}
Let us call the so-defined polytope $Pt_\lambda$ the 
{\it weight polytope}.  

If the weight $\lambda$ is regular, and the 
weight polytope is totally unimodular, then it is easy to see that 
the Brion formula gives
\begin{equation}
B_\lambda\ =\ 
\sum_{w\in W}\, e^{w\lambda}\, \prod_{\alpha\in S}\, 
(1-e^{-w\alpha})^{-1}\ ,  
\end{equation} 
since the feasible directions at the vertex $w\lambda$ are 
just the Weyl-transformed simple roots 
$\{ w\alpha \, |\, \alpha\in S \}$.

The last formula is remarkably similar to (3.5). We propose here to 
exploit this similarity, by expanding 
\begin{equation}
{\rm ch}_\lambda\ =\ \sum_{\mu\le\lambda}\, A_{\lambda,\mu}\, B_\mu\ .
\end{equation}
Here $\mu\le\lambda$ means $\lambda-\mu\in\mathbb Z_{\ge 0} R_>$, as usual. 
The constraint on the sum implies that the $A_{\lambda,\mu}$ are the 
entries of a triangular matrix. The 
character polytope-expansion (4.5) 
will manifest weight-multiplicity degeneracy 
beyond Weyl symmetry. 

Before considering examples, let us mention two possible complications. 
First, the weight $\lambda$ may 
not be regular, so that some subgroup of the Weyl group 
$W$ stabilizes it. We still find that the exponential sum for $Pt_\lambda$ 
equals $B_\lambda$, as written in (4.4). Also, when the algebra is not 
simply-laced, and $\lambda$ is not regular, then the corresponding weight 
polytope $Pt_\lambda$ may not be totally unimodular. 
Remarkably, even in that case, 
(4.4) seems to be the appropriate formula.  

\vskip10pt
\begin{quotation}
Consider a simple example, the adjoint representation of $G_2$. If 
$\alpha_2$ is the short simple root, then the highest weight $\lambda$ 
of the 
adjoint representation is $\Lambda^1 = \theta$. (Here  
$\theta$ will 
be used to denote the highest root.) The feasible directions at the 
vertex $\lambda=\Lambda^1$ are $-\alpha_1$ and $-\alpha_1-3\alpha_2$, 
{\it not} $-\alpha_1$ and $-\alpha_2$. The cone is not 
unimodular. Using (2.3), the cone function would be 
\begin{equation}
(1-e^{-\alpha_1})^{-1}(1-e^{-\alpha_1-3\alpha_2})^{-1}\ .
\end{equation}
This cone function would miss points -- in this example, 
the consequent formula 
for the polytope $Pt_\lambda$ would not have contributions from 
the short roots. 

As pointed out in \cite{B}, a cone can be decomposed into a 
set of unimodular ones, and (2.3) can then be applied. 
In general, however, it is more efficient to use a signed decomposition 
into unimodular cones. In this example, drawing a $G_2$ 
root diagram shows that 
\begin{equation}
\begin{split}
e^\lambda \sigma_\lambda\ =&\ e^{\Lambda^1}(1-e^{-\alpha_1})^{-1} 
(1-e^{-\alpha_2})^{-1} \\ 
\, &-\, e^{\Lambda^1-\alpha_2} 
(1-e^{-\alpha_1-3\alpha_2})^{-1}(1-e^{-\alpha_2})^{-1}\ .  
\end{split} 
\end{equation}
But this is just 
\begin{equation}
\begin{split}
e^\lambda \sigma_\lambda\ =&\ e^{\Lambda^1}(1-e^{-\alpha_1})^{-1} 
(1-e^{-\alpha_2})^{-1} \\ 
\, &+\, e^{r_2\Lambda^1} 
(1-e^{-r_2\alpha_1})^{-1}(-e^{-\alpha_2})(1-e^{-\alpha_2})^{-1}\\ 
=&\ e^{\Lambda^1}(1-e^{-\alpha_1})^{-1} 
(1-e^{-\alpha_2})^{-1} \\ 
\, &+\, e^{r_2\Lambda^1} 
(1-e^{-r_2\alpha_1})^{-1}(1-e^{-r_2\alpha_2})^{-1}\ . 
\end{split} 
\end{equation} 
By Weyl invariance, similar formulas work at the other vertices, 
and so (4.4) agrees with $E[Pt_\lambda; Q]$. 
\end{quotation}
\vskip10pt

We will proceed with our study of the expansion (4.5), postponing 
to \cite{GW} a 
proof that for all $\lambda\in P_\ge$, 
$E[Pt_\lambda; \lambda+Q] = B_\lambda$. Let us emphasize, 
however, that we are only relying on (4.4) and (4.5). Even if   
the sums of (4.3) and (4.4) are not identical, the expansion we are 
studying seems a natural one, and so should still be of value.

\vskip10pt
\begin{quotation}
Consider the simplest nontrivial case: $X_r=A_2$. Then 
$S=\{\alpha_1,\alpha_2\}$, and most importantly, 
$R_>\backslash S = \{\alpha_1+\alpha_2\} = \{\theta\}$. 
For $A_2$ then, we can write (4.4) as
\begin{equation}
\begin{split}
{\rm ch}_\lambda\ =&\ \sum_{w\in W} \, e^{w\lambda}\, 
\left[\prod_{\alpha\in S}(1-e^{-w\alpha})^{-1}\right]\, 
(1-e^{-w\theta})^{-1}\ \\ 
=&\ B_\lambda + \sum_{w\in W} \, e^{w(\lambda-\theta)}\, 
\left[\prod_{\alpha\in S}(1-e^{-w\alpha})^{-1}\right]\, 
(1-e^{-w\theta})^{-1}\ ,
\end{split}
\end{equation}
so that 
\begin{equation}
{\rm ch}_\lambda\ =\ B_\lambda\ +\ {\rm ch}_{\lambda-\theta}\ .
\end{equation}
It is also simple to show that if either $\nu_1=0$ or $\nu_2=0$, then 
${\rm ch}_\nu=B_\nu$. Since $\theta=\Lambda^1+\Lambda^2$ for $A_2$, 
we find that if $\lambda_{\rm min}:=\min\{\lambda_1,\lambda_2\}$, then 
\begin{equation}
{\rm ch}_\lambda\ =\ B_\lambda\ +\ B_{\lambda-\theta}\ +\ 
B_{\lambda-2\theta}\ +\ \ldots\ +\ B_{\lambda-\lambda_{\rm min}\theta}\ .
\end{equation}
Eqn. (4.11) manifests the weight multiplicity pattern of $A_2$ 
representations.\footnote{\,This formula was derived in 
\cite{AS}, where weight multiplicity patterns were studied for low-rank 
algebras, using methods close in spirit to the Kostant multiplicity 
formula \cite{K}. The $A_2$ multiplicity pattern was known long 
before, however, by Wigner, for example; I thank Professor R. King 
for so informing me.}

It is difficult to generalize the derivation just given of (4.11) 
to other algebras. (4.10) is easier, however. Consider $X_r=C_2$, with 
the short simple root labelled as $\alpha_1$. Then 
\begin{equation}
R_>\backslash S\ =\ \{\, 2\alpha_1+\alpha_2=2\Lambda^1,\, 
                     \alpha_1+\alpha_2=\Lambda^2 \,\}\ .
\end{equation}
Therefore 
\begin{equation}
\begin{split}
{\rm ch}_\lambda\ =&\ \sum_{w\in W} \, e^{w\lambda}\, 
\left[\prod_{\alpha\in S}(1-e^{-w\alpha})^{-1}\right]\, 
(1-e^{-w(2\Lambda^1)})^{-1}(1-e^{-w\Lambda^2})^{-1}\ \\ 
=&\ B_\lambda + \sum_{w\in W} \, e^{w\lambda}\, 
\left(e^{w(-2\Lambda^1)}+e^{w(-\Lambda^2)}-e^{w(-2\Lambda^1+\Lambda^2)}
\right)\,\\ & \qquad\qquad\qquad\qquad\qquad\qquad\times\, 
\left[\prod_{\alpha\in R_>}(1-e^{-w\alpha})^{-1}\right]\ ,
\end{split}
\end{equation}
so that 
\begin{equation}
{\rm ch}_\lambda\ =\ B_\lambda\ +\ {\rm ch}_{\lambda-2\Lambda^1}\ +\ 
{\rm ch}_{\lambda-\Lambda^2}\ -\ {\rm ch}_{\lambda-2\Lambda^1-\Lambda^2}\ .
\end{equation}

This recurrence relation is a bit more complicated 
than that of $A_2$, but can 
be analysed easily. We need the relation 
\begin{equation}
{\rm ch}_\lambda\ =\ (\det w)\, {\rm ch}_{w.\lambda}\ ,
\end{equation}
derived from either (3.2) or (3.3). Using it with (4.14), we can establish 
\begin{equation}
{\rm ch}_{\lambda_j\Lambda^j}\ =\ B_{\lambda_j\Lambda^j}\ +\ 
{\rm ch}_{(\lambda_j-2)\Lambda^j}\ ,\ \ \ \ \ (j=1,2).
\end{equation}
This immediately shows that ${\rm ch}_\lambda=B_\lambda$ for 
$\lambda\in\{0,\Lambda^1,\Lambda^2\}$, and that 
\begin{equation}
{\rm ch}_{\lambda_j\Lambda^j}\ =\ B_{\lambda_j\Lambda^j} + 
B_{(\lambda_j-2)\Lambda^j} +  \ldots +
B_{[\lambda_j]_2\Lambda^j}
\ ,\ \ (j=1,2). 
\end{equation} 
Here $[\lambda_j]_2:=0\,(1)$ if $\lambda_j$ is even (odd). Now, if we define 
\begin{equation}
v_\lambda\ :=\ {\rm ch}_\lambda\ -\ {\rm ch}_{\lambda-2\Lambda^1}\ ,
\end{equation}
then the recursion relation (4.14) becomes 
\begin{equation}
\begin{split}
v_\lambda\ =&\ B_\lambda\ +\ v_{\lambda-\Lambda^2} \\ 
=&\ B_\lambda\ +\ B_{\lambda-\Lambda^2}\ +\ \ldots\ +\ 
B_{\lambda_1\Lambda^1}\ , 
\end{split}
\end{equation}
since (4.17) implies $v_{\lambda_1\Lambda^1}=
B_{\lambda_1\Lambda^1}.$ Solving (4.18) yields 
\begin{equation}
{\rm ch}_\lambda\ =\ v_\lambda + v_{\lambda-2\Lambda^1} + \ldots 
+ v_{\Lambda^1+\lambda_2\Lambda^2} 
\end{equation}
for $\lambda_1$ odd; and 
\begin{equation}
{\rm ch}_\lambda\ =\ v_\lambda + v_{\lambda-2\Lambda^1} + \ldots 
+ v_{2\Lambda^1+\lambda_2\Lambda^2}+{\rm ch}_{\lambda_2\Lambda^2}
\end{equation}
for $\lambda_1$ even. 
Using first (4.20), then (4.19), we find 
\begin{equation}
\begin{split}
{\rm ch}_\lambda\ =&\ B_\lambda + B_{\lambda-\Lambda^2} + \ldots 
+ B_{\lambda_1\Lambda^1}\\ 
&+ B_{\lambda-2\Lambda^1} + B_{\lambda-2\Lambda^1-\Lambda^2} + \ldots 
+ B_{(\lambda_1-2)\Lambda^1}\\ 
&+ \ldots \\ 
&+ B_{\Lambda^1+\lambda_2\Lambda^2} + B_{\Lambda^1+(\lambda_2-1)\Lambda} 
+ \ldots 
+ B_{\Lambda^1}\\ 
\end{split}
\end{equation}
for $\lambda_1$ odd. Replacing (4.20) by (4.21) and (4.17), we get instead 
\begin{equation}
\begin{split}
{\rm ch}_\lambda\ =&\ B_\lambda + B_{\lambda-\Lambda^2} + \ldots 
+ B_{\lambda_1\Lambda^1}\\ 
&+ B_{\lambda-2\Lambda^1} + B_{\lambda-2\Lambda^1-\Lambda^2} + \ldots 
+ B_{(\lambda_1-2)\Lambda^1}\\ 
&+ \ldots \\ 
&+ B_{\lambda_2\Lambda^2} + B_{(\lambda_2-2)\Lambda^2} +\ldots + 
B_{[\Lambda_2]_2\Lambda^2}\\ 
\end{split}
\end{equation}
for $\lambda_1$ even. These $C_2$ results confirm those of \cite{AS}.
\end{quotation}
\vskip10pt 

Re-writing (4.10) as $B_\lambda = {\rm ch}_\lambda - 
{\rm ch}_{\lambda-\theta}$ gives us a hint as to how to generalize 
the method of computation of the $A_{\lambda,\mu}$ 
to all algebras. Expanding  
\begin{equation}
B_\lambda\ =\ \sum_{\mu\le \lambda}\, A^{-1}_{\lambda,\mu}\, 
{\rm ch}_\mu\ , 
\end{equation}
is straightforward. Then finding $A_{\lambda,\mu}$ by diagonalizing the 
triangular matrix $(A^{-1}_{\lambda,\mu})$ is relatively 
easy.\footnote{\,The author learned this trick from a work of 
Professors J. Patera and  R. T. Sharp \cite{PS1}, and has 
also used it elsewhere \cite{GJW}.}

First, re-write (4.4) as 
\begin{equation}
B_\lambda\ =\ \sum_{w\in W}\, e^{w\lambda}\, 
\left[\prod_{\gamma\in R_>\backslash S}(1-e^{-w\gamma})\right]\,  
\left[\prod_{\alpha\in R_>}(1-e^{-w\alpha})^{-1}\right]\ . 
\end{equation}
Comparing to (3.5), we can therefore write 
\begin{equation}
B_\lambda\ =\ \widehat{\rm ch}\ e^\lambda\, 
\prod_{\gamma\in R_>\backslash S}\, (1-e^{-\gamma})\ ,
\end{equation}
where we have defined 
\begin{equation}
\widehat{\rm ch}\, e^\lambda\ :=\ {\rm ch}_\lambda\ .
\end{equation}
Of course, some of the terms of the expansion just written may be of 
the form ${\rm ch}_\lambda$, but with $\lambda\not\in P_\ge$. 
To find the coefficients $A^{-1}_{\lambda,\mu}$ with 
$\lambda,\mu\in P_\ge$, therefore, it is  
necessary to use the relation (4.15).

To write an explicit formula for $A^{-1}_{\lambda,\mu}$, we first 
define a partition function $F$ by
\begin{equation}
\prod_{\gamma\in R_>\backslash S}\, (1-e^{-\gamma})\ =:\ 
\sum_{\beta\in \mathbb Z_{\ge 0} R_>}\, F(\beta)\, e^\beta\ .
\end{equation} 
Then 
\begin{equation}
B_\lambda\ =\ \sum_{\beta\in \mathbb Z_{\ge 0} R_>}\, F(\beta)\, 
{\rm ch}_{\lambda-\beta}\ . 
\end{equation}
Using (4.24) and (4.15), we find
\begin{equation}
A^{-1}_{\lambda,\mu}\ =\ \sum_{w\in W}\, (\det w)\, F(\lambda-w.\mu)\ .
\end{equation}

\vskip10pt
\begin{quotation}
For $X_r=G_2$, we have
\begin{equation}
\begin{split}
R_>\backslash S\ =&\ \{\, \alpha_1+\alpha_2,\, 2\alpha_1+3\alpha_2,\, 
\alpha_1+2\alpha_2,\, \alpha_1+3\alpha_2\, \}\ , \\ 
=&\ \{\, \Lambda^1-\Lambda^2,\, \Lambda^1,\, 
\Lambda^2,\, -\Lambda^1+3\Lambda^2\, \}\ .   
\end{split}
\end{equation}
Therefore (4.26) yields
\begin{equation}
\begin{split}
B_\lambda\ =&\ {\rm ch}_{\lambda} - 
{\rm ch}_{\lambda+\Lambda^1-3\Lambda^2} -
{\rm ch}_{\lambda-\Lambda^2} + 
{\rm ch}_{\lambda+\Lambda^1-4\Lambda^2}  \\  
& + {\rm ch}_{\lambda-\Lambda^1-\Lambda^2} - 
{\rm ch}_{\lambda-4\Lambda^2} - 
{\rm ch}_{\lambda-\Lambda^1+\Lambda^2}  \\ 
& + {\rm ch}_{\lambda-2\Lambda^2} + 
{\rm ch}_{\lambda-2\Lambda^1+\Lambda^2}  \\ 
& - {\rm ch}_{\lambda-\Lambda^1-2\Lambda^2} - 
{\rm ch}_{\lambda-2\Lambda^1} + 
{\rm ch}_{\lambda-\Lambda^1-3\Lambda^2}\ .
\end{split}
\end{equation}
A simple calculation then yields the matrix $(A_{\lambda,\mu})$:
\begin{equation}
\left [\begin {array}{cccccccccccccccc} 1&0&0&0&0&0&0&0&0&0&0&0&0&0&0&0
\\\noalign{\medskip}0&1&0&0&0&0&0&0&0&0&0&0&0&0&0&0
\\\noalign{\medskip}1&0&1&0&0&0&0&0&0&0&0&0&0&0&0&0
\\\noalign{\medskip}1&1&0&1&0&0&0&0&0&0&0&0&0&0&0&0
\\\noalign{\medskip}0&2&0&1&1&0&0&0&0&0&0&0&0&0&0&0
\\\noalign{\medskip}2&1&1&1&0&1&0&0&0&0&0&0&0&0&0&0
\\\noalign{\medskip}1&2&0&2&1&0&1&0&0&0&0&0&0&0&0&0
\\\noalign{\medskip}2&2&1&2&1&1&0&1&0&0&0&0&0&0&0&0
\\\noalign{\medskip}2&2&0&2&1&1&0&0&1&0&0&0&0&0&0&0
\\\noalign{\medskip}1&3&0&2&2&1&0&1&0&1&0&0&0&0&0&0
\\\noalign{\medskip}3&4&1&4&2&2&1&1&1&0&1&0&0&0&0&0
\\\noalign{\medskip}3&2&1&2&1&2&0&1&1&0&0&1&0&0&0&0
\\\noalign{\medskip}2&2&1&2&1&1&1&1&0&0&0&0&1&0&0&0
\\\noalign{\medskip}3&5&0&5&3&2&1&2&1&1&1&0&0&1&0&0
\\\noalign{\medskip}4&4&1&4&3&3&0&2&2&1&1&1&0&0&1&0
\\\noalign{\medskip}3&4&0&4&2&2&1&1&2&0&1&1&0&0&0&1\end {array}\right
]
\end{equation}
Here all weights $\lambda=:(\lambda_1,\lambda_2)$ 
with $\lambda\cdot\theta = 2\lambda_1+\lambda_2\le 6$ are 
included, in the order
\begin{equation}
\begin{split}
&(0,0),\,(0,1),\,(1,0),\,(0,2),\,(1,1),\,(0,3),\,
(2,0),\,(1,2),\\ 
&(0,4),\,(2,1),\,(1,3),\,(0,5),\,
(3,0),\,(2,2),\,(1,4),\,(0,6)\ .
\end{split}
\end{equation}
\end{quotation}
\vskip10pt

The character polytope expansion (4.5) can be combined with the 
Weyl dimension formula for the dimension $d_\lambda$ 
of the representation of 
highest weight $\lambda$:
\begin{equation}
d_\lambda\ =\ \prod_{\alpha\in R_>}\, 
\frac{(\lambda+\rho)\cdot\alpha }{\rho\cdot\alpha }\ .
\end{equation} 
The result is a formula for the number $b_\lambda$ 
of lattice points counted by 
$B_\lambda$, that provides helpful checks on any expansions derived, since
(4.24) and (4.5) imply $b_\lambda = \sum_\mu A^{-1}_{\lambda,\mu} d_\mu$ and 
$d_\mu = \sum_\sigma A_{\mu,\sigma} b_\sigma$, respectively. 
The formulas relevant to the simple rank-two algebras are
\begin{equation}
\begin{split}
A_2:\qquad\qquad &b_\lambda\ =\ (\lambda_1^2+4\lambda_1\lambda_2+ 
\lambda_2^2+ 3\lambda_1+3\lambda_2 +2)/2\ , \\ 
C_2:\qquad\qquad &b_\lambda\ =\ \lambda_1^2+4\lambda_1\lambda_2+ 
2\lambda_2^2+ 2\lambda_1+2\lambda_2 +1\ , \\ 
G_2:\qquad\qquad &b_\lambda\ =\ 9\lambda_1^2+12\lambda_1\lambda_2+ 
3\lambda_2^2+ 3\lambda_1+3\lambda_2 + 1\ .  
\end{split}
\end{equation}

\vskip10pt
\begin{quotation}
As our final example, consider $X_r=A_3$. The important subset of 
positive roots is 
\begin{equation}
\begin{split}
R_>\backslash S\ =&\ \{\, \alpha_{12},\, 
\alpha_{123},\,
\alpha_{23}
\,\}\\ 
=&\ \{\, \Lambda^1+\Lambda^2-\Lambda^3,\, 
\Lambda^1+\Lambda^3,\,
-\Lambda^1+\Lambda^2+\Lambda^3
\,\}\ ,
\end{split}
\end{equation}
where $\alpha_{12}:=\alpha_1+\alpha_2$, etc. We therefore find 
\begin{equation}
\begin{split}
B_\lambda\ =&\ {\rm ch}_{\lambda} - 
{\rm ch}_{\lambda+\Lambda^1-\Lambda^2-\Lambda^3} -
{\rm ch}_{\lambda-\Lambda^1-\Lambda^3} + 
{\rm ch}_{\lambda-\Lambda^2-2\Lambda^3}  \\  
&- {\rm ch}_{\lambda-\Lambda^1-\Lambda^2+\Lambda^3} + 
{\rm ch}_{\lambda-2\Lambda^2} + 
{\rm ch}_{\lambda-2\Lambda^1-\Lambda^2}  \\ 
& - {\rm ch}_{\lambda-\Lambda^1-2\Lambda^2-\Lambda^3}\ .
\end{split}
\end{equation}
Using the Weyl dimension formula, 
\begin{equation}
\begin{split}
b_\lambda\ =&\ 1 + (11\lambda_1+14\lambda_2+11\lambda_3)/6\\
&+ 4\lambda_1\lambda_2 +3\lambda_1\lambda_3+4\lambda_2\lambda_3 
+\lambda_1^2 +2\lambda_2^2 +\lambda_3^2 \\
&+ (36\lambda_1\lambda_2\lambda_3 
+12\lambda_1\lambda_2^2 +12\lambda_2^2\lambda_3 
+6\lambda_1^2\lambda_2 +6\lambda_2\lambda_3^2\\ 
&+9\lambda_1^2\lambda_3 +9\lambda_1\lambda_3^2  
+\lambda_1^3 +4\lambda_2^3 + \lambda_3^2)/6
\end{split}
\end{equation}
follows.

We present our results for all weights $\lambda
=:(\lambda_1,\lambda_2,\lambda_3)\in P_\ge$ of 
$A_3$, with $\lambda\cdot\theta = \lambda_1+\lambda_2+\lambda_3 
\le 5$. The weights separate into four congruence classes, corresponding to 
the four shifted root lattices $\{0,\Lambda^1,\Lambda^2,\Lambda^3\}
+Q$ that combine to form the weight lattice $P$.   

For weights $\lambda,\mu\in Q$, we find the matrix 
$(A_{\lambda,\mu})$ is
\begin{equation}
\left [\begin {array}{cccccccccccccc} 1&0&0&0&0&0&0&0&0&0&0&0&0&0
\\\noalign{\medskip}2&1&0&0&0&0&0&0&0&0&0&0&0&0\\\noalign{\medskip}1&0
&1&0&0&0&0&0&0&0&0&0&0&0\\\noalign{\medskip}1&1&0&1&0&0&0&0&0&0&0&0&0&0
\\\noalign{\medskip}1&1&0&0&1&0&0&0&0&0&0&0&0&0\\\noalign{\medskip}0&0
&0&0&0&1&0&0&0&0&0&0&0&0\\\noalign{\medskip}1&0&1&0&0&0&1&0&0&0&0&0&0&0
\\\noalign{\medskip}0&0&0&0&0&0&0&1&0&0&0&0&0&0\\\noalign{\medskip}2&1
&1&1&1&0&0&0&1&0&0&0&0&0\\\noalign{\medskip}3&2&0&0&0&0&0&0&0&1&0&0&0&0
\\\noalign{\medskip}2&2&0&1&0&1&0&0&0&1&1&0&0&0\\\noalign{\medskip}2&2
&0&0&1&0&0&1&0&1&0&1&0&0\\\noalign{\medskip}1&1&0&1&1&0&0&0&1&0&0&0&1&0
\\\noalign{\medskip}1&1&0&1&1&0&0&0&1&0&0&0&0&1\end {array}\right ]\ .
\end{equation}
In the order used, the weights are 
\begin{equation}
\begin{split}
&(0,0,0),\ (1,0,1),\ (0,2,0),\ (2,1,0),\ (0,1,2),\ (4,0,0),\ (0,4,0),\ \\ 
&(0,0,4),\ (1,2,1),\ (2,0,2),\ (3,1,1),\ (1,1,3),\ (2,3,0),\ (0,3,2)\ .
\end{split}
\end{equation}

For weights $\lambda,\mu\in \Lambda^1+Q$, we find 
$(A_{\lambda,\mu})$ to be  
\begin{equation}
\left [\begin {array}{cccccccccccccc} 1&0&0&0&0&0&0&0&0&0&0&0&0&0
\\\noalign{\medskip}1&1&0&0&0&0&0&0&0&0&0&0&0&0\\\noalign{\medskip}0&0
&1&0&0&0&0&0&0&0&0&0&0&0\\\noalign{\medskip}1&1&0&1&0&0&0&0&0&0&0&0&0&0
\\\noalign{\medskip}2&0&0&0&1&0&0&0&0&0&0&0&0&0\\\noalign{\medskip}1&1
&0&1&0&1&0&0&0&0&0&0&0&0\\\noalign{\medskip}2&1&1&0&1&0&1&0&0&0&0&0&0&0
\\\noalign{\medskip}1&0&0&0&1&0&0&1&0&0&0&0&0&0\\\noalign{\medskip}1&0
&1&0&1&0&1&0&1&0&0&0&0&0\\\noalign{\medskip}0&0&2&0&0&0&0&0&0&1&0&0&0&0
\\\noalign{\medskip}1&1&0&1&0&1&0&0&0&0&1&0&0&0\\\noalign{\medskip}2&1
&1&1&1&0&1&1&0&0&0&1&0&0\\\noalign{\medskip}3&0&0&0&2&0&0&0&0&0&0&0&1&0
\\\noalign{\medskip}0&0&0&0&0&0&0&0&0&0&0&0&0&1\end {array}\right ]
\end{equation}
for weights $(\lambda_1,\lambda_2,\lambda_3)$ in the order
\begin{equation}
\begin{split}
&(1,0,0),\ (0,1,1),\ (0,0,3),\ (1,2,0),\ (2,0,1),\ (0,3,1),\ (1,1,2),\ \\ 
&(3,1,0),\ (0,2,3),\ (1,0,4),\ (1,4,0),\ (2,2,1),\ (3,0,2),\ (5,0,0)\ .
\end{split}
\end{equation}
The results for weights in $\Lambda^3+Q$ can be obtained from this, by 
replacing all weights $(\lambda_1,\lambda_2,\lambda_3)$ by their charge 
conjugates $(\lambda_3,\lambda_2,\lambda_1)$. 

The matrix 
$(A_{\lambda,\mu})$ is
\begin{equation} 
\left [\begin {array}{cccccccccccccc} 1&0&0&0&0&0&0&0&0&0&0&0&0&0
\\\noalign{\medskip}0&1&0&0&0&0&0&0&0&0&0&0&0&0\\\noalign{\medskip}0&0
&1&0&0&0&0&0&0&0&0&0&0&0\\\noalign{\medskip}1&0&0&1&0&0&0&0&0&0&0&0&0&0
\\\noalign{\medskip}1&1&1&0&1&0&0&0&0&0&0&0&0&0\\\noalign{\medskip}0&2
&0&0&0&1&0&0&0&0&0&0&0&0\\\noalign{\medskip}0&0&2&0&0&0&1&0&0&0&0&0&0&0
\\\noalign{\medskip}0&1&1&0&1&0&0&1&0&0&0&0&0&0\\\noalign{\medskip}0&1
&1&0&1&0&0&0&1&0&0&0&0&0\\\noalign{\medskip}0&1&0&0&0&1&0&0&0&1&0&0&0&0
\\\noalign{\medskip}1&0&0&1&0&0&0&0&0&0&1&0&0&0\\\noalign{\medskip}1&1
&1&1&1&0&0&1&1&0&0&1&0&0\\\noalign{\medskip}1&2&2&0&1&1&1&0&0&0&0&0&1&0
\\\noalign{\medskip}0&0&1&0&0&0&1&0&0&0&0&0&0&1\end {array}\right ]
\end{equation}
for weights $\lambda,\mu\in \Lambda^2+Q$. The order of weights is 
\begin{equation}
\begin{split}
&(0,1,0),\ (0,0,2),\ (2,0,0),\ (0,3,0),\ (1,1,1),\ (1,0,3),\ (3,0,1),\ \\ 
&(0,2,2),\ (2,2,0),\ (0,1,4),\ (0,5,0),\ (1,3,1),\ (2,1,2),\ (4,1,0)\ .
\end{split}
\end{equation}
\end{quotation}
\vskip10pt

The preliminary calculations we have done for $G_2$ and $A_3$ 
confirm that the polytope-expansion multiplicities $A_{\lambda,\mu}$ can 
be computed easily by computer. Some patterns can already be seen from 
their results. To make some general statements, however, we plan 
to attempt an analysis of the recursion relations modeled on the one done 
above for $C_2$ \cite{GW}. Perhaps a relatively simple algorithm, of a 
combinatorial type, can be found for the calculation of the 
$A_{\lambda,\mu}$.

\vskip20pt
\section{Conclusion} 

First, we'll summarize the main points. Then we'll discuss what still needs 
to be done, and what might be done. 

We point out that the Brion formula applied to the weight polytopes 
$Pt_\lambda$ produces a formula (4.4) that is remarkably similar to 
the Weyl character formula (3.5). The polytope expansion (4.5) of the 
Lie characters was therefore advocated. It 
makes manifest certain degeneracies in weight multiplicities beyond those 
explained by Weyl group symmetry, for example. The expansion 
multiplicities $A_{\lambda,\mu}$ were studied. A closed formula 
(4.30) was given for them, and methods for their computation 
were discussed and illustrated with the examples of $A_2,C_2,G_2$ and 
$A_3$.  

Two conjectures need to be established. A general proof that the 
polytope-expansion coefficients are non-negative integers, 
\begin{equation}
A_{\lambda,\mu}\ \in\  \mathbb Z_{\ge 0}\ ,
\end{equation} 
would be helpful. It should also be determined if 
the exponential sum $E[Pt_\lambda; \lambda+Q]$ of the 
weight polytope $Pt_\lambda$ 
is given by the Brion expression $B_\lambda$ in (4.4), i.e., if 
\begin{equation}
\sum_{\mu\in P(\lambda)}\,e^\mu\ =\ 
\sum_{w\in W}\, e^{w\lambda}\, \prod_{\alpha\in S}\, 
(1-e^{-w\alpha})^{-1}\ , 
\end{equation}
for all integrable highest weights $\lambda\in P_\ge$. As mentioned above, 
however, we use only (4.4) and (4.5). Our results therefore have value 
even if this last equality is not always obeyed. 

We'll now mention a few other possible applications of 
polytope theory to Lie theory. 

This author was introduced to polytope theory in different contexts --  
in the study of tensor product multiplicities and the related affine fusion 
multiplicities (see \cite{RW} and references therein, and \cite{BCMW}), 
the fusion of 
Wess-Zumino-Witten conformal field theories. It 
would be interesting to consider applications of the Brion formula in   
those subjects. 

For example, the Brion formula also allows the derivation of a character 
generating function relevant to affine fusion. Its derivation is a 
simple adaptation of that of the Patera-Sharp formula for the 
character generator of a simple Lie algebra \cite{PS2}. We hope to report 
on it and its applications elsewhere.

As another example, consider the tensor-product multiplicity patterns for 
$A_2$ that were studied thoroughly long ago, in \cite{PSV} and \cite{PP}. 
Not surprisingly, the pattern is similar to the weight-multiplicity pattern 
of a single irreducible $A_2$ representation. Perhaps a polytope expansion 
in the spirit of (4.5) could manifest properties of the tensor-product 
patterns for all simple Lie algebras. 

Recent work on tensor products has provided results 
that are valid for general classes of algebras, but on the 
simpler question of which weights appear in a given tensor product, 
ignoring their multiplicities. See \cite{F} for a review. The Brion 
formula might be useful for this question, and perhaps also for 
the corresponding problem in affine fusion \cite{AW}. 

Finally, it is our hope that applications 
in the other direction will also be found, i.e., that 
certain techniques from Lie theory can also help 
in the study of polytopes. For example, the Brion formula has  
similarities with the Weyl character formula, as we have discussed. Are 
there polytope formulas that correspond to other formulas for Lie 
characters?

\vskip30pt
\bibliographystyle{amsplain}

\begin{thebibliography}{10} 

\bibitem {AW}
S. Agnihotri, C. Woodward, 
\textit{Eigenvalues of products of unitary matrices and 
quantum Schubert calculus},  
Math. Res. Lett. \textbf{5} (1998), no. 6, 817--836.


\bibitem {AS} 
J.-P. Antoine, D. Speiser,  
\textit{Characters of irreducible representations of the simple groups. 
I. General theory.}  
J. Mathematical Phys. {\bf 5} (1964) 1226--1234; \hfill\break 
\textit{Characters of irreducible representations of the simple groups. II. 
Application to classical groups},  
J. Mathematical Phys. {\bf 5} (1964) 1560--1572.

\bibitem {B} A. Barvinok, 
\textit{Lattice points and lattice polytopes}.
Handbook of discrete and computational geometry, 133--152,
CRC Press Ser. Discrete Math. Appl.,
(CRC, Boca Raton, FL, 1997). 

\bibitem {BCMW} 
L. B\'egin, C. Cummins, P. Mathieu, M.A. Walton, 
\textit{Fusion rules and the Patera-Sharp generating-function method}, 
these proceedings (hep-th/0210182). 

\bibitem {Br} M. Brion, 
\textit{Poly\`edres et r\'eseaux}, 
Enseign. Math. (2) \textbf{38} (1992), no. 1-2, 71--88. 

\bibitem{F}
W. Fulton, 
\textit{Eigenvalues, invariant factors, highest 
weights, and Schubert calculus},  
Bull. Amer. Math. Soc. (N.S.) \textbf{37} (2000), no. 3, 209--249 (electronic).

\bibitem {GJW} T. Gannon, C. Jakovljevic, M. A. Walton, 
\textit{Lie group weight multiplicities from conformal field theory}, 
J. Phys. A \textbf{28} (1995), no. 9, 2617--2625.

\bibitem {GW} T. Gannon, M. A. Walton, in preparation. 

\bibitem {K} B. Kostant, 
\textit{A formula for the multiplicity of a weight}, 
Trans. Amer. Math. Soc. \textbf{93} (1959) 53--73. 

\bibitem {PP} A. M. Perelomov, V. S. Popov, 
\textit{Expansion of the direct product of 
irreducible representations of ${\rm SU}(3)$ in irreducible representations},  
Jadernaja Fiz. \textbf{2} (1965) 294--306 (Russian); 
translated as Soviet J. Nuclear Phys. \textbf{2} (1965) 210--218.

\bibitem {PSV} 
B. Preziosi, A. Simoni, B. Vitale,  
\textit{A general analysis of the reduction of the 
direct product of two irreducible representations of 
${\rm SU}\sb{3}$ and of its multiplicity structure}, 
Nuovo Cimento (10) \textbf{34} (1964), no. 10, 1101--1113.

\bibitem {PS1} 
J. Patera, R. T. Sharp,  
\textit{Branching rules for representations of 
simple Lie algebras through Weyl group orbit reduction},  
J. Phys. A \textbf{22} (1989), no. 13, 2329--2340.

\bibitem {PS2} 
J. Patera, R. T. Sharp,  
\textit{Generating function techniques pertinent to 
spectroscopy and crystal physics}, in  
Recent advances in group theory and their application to spectroscopy, 
219--248,  
NATO Adv. Study Inst. Ser., Ser. B: Physics, 43, 
(Plenum, New York-London, 1979). 

\bibitem {RW} 
J. Rasmussen, M. A. Walton, 
\textit{${su}(N)$ tensor product multiplicities and virtual 
Berenstein-Zelevinsky triangles}, 
J. Phys. A \textbf{34} (2001), no. 49, 11095--11105;\hfill\break 
\textit{Affine ${su}(3)$ and ${su}(4)$ fusion multiplicities as 
polytope volumes}, J. Phys. A \textbf{35} (2002), no. 32, 6939-6952. 

\bibitem {Zi} G. Ziegler, 
\textit{Lectures on polytopes}, 
Graduate Texts in Mathematics, 152
(Springer-Verlag, New York, 1995).








\end{thebibliography}

\end{document}